\documentclass[preprint,authoryear,12pt]{article}

\usepackage{amsmath}
\usepackage{amssymb}
\usepackage{multirow}
\usepackage[english]{babel}
\usepackage{graphicx}
\usepackage{epstopdf}
\usepackage{color}
\usepackage{natbib}
\usepackage{amssymb}
\usepackage[margin=0.6 in]{geometry}
\usepackage{setspace}
\usepackage{enumerate}

\renewcommand{\vec}[1]{\mathbf{#1}}
\newcommand{\svec}[1]{\boldsymbol{#1}}

\newcommand{\yv}{\vec{y}}

\newcommand{\yb}{\bar{y}}

\newcommand{\etav}{\svec{\eta}}

\newcommand{\Bh}{\hat{B}}
\newcommand{\BMDh}{\widehat{BMD}}

\newcommand{\betav}{\svec{\beta}}

\newcommand{\betavh}{\hat{\svec{\beta}}}

\newcommand{\alphav}{\svec{\alpha}}
\newcommand{\deltav}{\svec{\delta}}
\newcommand{\Lambdav}{\svec{\Lambda}}
\newcommand{\Omegav}{\svec{\Omega}}

\newcommand{\Hv}{\vec{H}}

\newcommand{\etah}{\hat{\etav}}

\newcommand{\deltavh}{\hat{\deltav}}
\newcommand{\alphavh}{\hat{\alphav}}

\newcommand{\sigmah}{\hat{\sigma}}

\newcommand{\Jv}{\vec{J}}

\numberwithin{equation}{section}

\begin{document}

\title{Benchmark Dose Estimation using a Family of Link Functions}

\author{ I.~Das\thanks{{Corresponding author. Email}:
id31@duke.edu, \vspace{6pt} {Phone}: +19195192053}\\
Department of Statistical Science, Duke University, NC, USA}



\date{}
\maketitle
\begin{abstract}
This article proposes a method of estimating benchmark dose (BMD) using a family of link functions in binomial response models dealing with model uncertainty problems. Researchers usually estimate the BMD using binomial response models with a single link function. Several forms of link function have been proposed to fit dose response models to estimate the BMD and the corresponding benchmark dose lower bound (BMDL). However, if the assumed link is not correct, then the estimated BMD and BMDL from the fitted model may not be accurate. To account for model uncertainty, model averaging (MA) methods are proposed to estimate BMD averaging over a model space containing a finite number of standard models. Usual model averaging focuses on a pre-specified list of parametric models leading to pitfalls when none of the models in the list is the correct model. Here, an alternative which augments an initial list of parametric models with an infinite number of additional models having varying links has been proposed. In addition, different methods for estimating BMDL based on the family of link functions are derived. The proposed approach is compared with MA in a simulation study and applied to a real data set.  Simulation studies are also conducted to compare the four methods of estimating BMDL.
\end{abstract}

\newpage
\section{Introduction}
One of the main goals in quantitative risk assessment is to estimate the risk function $R(d)$, which is the probability of adverse events, such as death, birth defect, weight loss, cancer or mutation exhibited in a subject
exposed at dose level $d$. Suppose $n$ number of subjects are exposed to a dose level $d$ and $y$ number of adverse events are observed. Then, the response $y$ is distributed according to a binomial distribution with parameter $[n,R(d)]$, where $R(d)$ is the probability of adverse events at dose level $d$. After estimating the risk function $R(d)$, the extra risk function $R_E(d)$, defined as $R_E(d)=\frac{R(d)-p_0}{1-p_0}$ is computed, where $p_0$ is the risk at minimum dose level usually called as background risk. The benchmark dose (BMD) is defined by the dose level having the extra risk $R_E(BMD)=BMR$, where $BMR$ is called the benchmark response usually pre-specified as $0.01,0.05$, or $0.1$. The benchmark dose lower bound (BMDL) is also determined using the risk function $R(d)$. The accuracy of the estimation of BMD and BMDL is dependent upon the estimation of the risk function $R(d)$.\par

Methods of estimating BMD and BMDL are discussed by several researchers such as \cite{1984_crump,2005_bailer,morales2006bayesian,wheeler2007properties,wheeler2009comparing,west2012impact} to name just a few. \cite{1984_crump} introduced methods of estimating BMD and BMDL by proposing four models for discrete responses and three models for continuous responses. There are eight models \citep{wheeler2007properties,wheeler2009comparing,west2012impact} that have been identified as standard models for estimating BMD and BMDL.  One of the models from the set of standard models may be chosen for fitting the data sets. However, the responses may be generated from the model other than the chosen model. Researchers \citep{wheeler2007properties,wheeler2009comparing,west2012impact} have shown that the estimation of BMD and BMDL are significantly effected if the assumed model is incorrect. So, there is a recent rise in developing methods of accounting for model uncertainty in BMD estimation. \par
 For accounting model uncertainty in BMD and BMDL estimation, model averaging (MA) methods are proposed by \cite{kang2000incorporating,2005_bailer,wheeler2007properties,shao2011potential,west2012impact,piegorsch2013information}. The estimates of BMD and BMDL using model averaging methods are given by the weighted average of the estimates of BMD and BMDL using individual models belong to a set of models. Bayesian methods and Bayesian model averaging methods for estimating BMD and BMDL are also proposed by \cite{morales2006bayesian,shao2012statistical,2015_simmons}. The model averaging approach may solve the problems of model uncertainty, when the true model generating responses can be approximated by some members of the model space containing the assumed models. Since, the model space are always finite, there may be infinite number of other models which can not be approximated by the members of the model space. So, model averaging techniques provide a partial solution to the problem of model uncertainty.  \par

Here, a family of link functions containing some of the standard link functions as well as infinite number of other link functions is used to fit the binomial response models. The family of link functions is parameterized by two unknown link parameters. There are infinite number of link functions can be represented by different values of link parameters. Some standard link functions correspond to some finite values of link parameters. So, we may get a better results for accounting model uncertainty in BMD and BMDL estimation using the proposed model.\par

The remainder of the article is organized as follows: in Section \ref{sm}, the binomial response models using a family of link functions are discussed. An expression for $BMD$ using the family of link functions is given in Section \ref{sbmd}. An example with real data set is shown in Sections \ref{ser} to illustrate the proposed method of estimating BMD. Four methods of estimating BMDL are derived in Section \ref{sbmdl} and a comparison study among the four methods are provided in Section \ref{scs}. In Section \ref{scompflma}, the proposed method is compared with model averaging method using simulation studies. Concluding remarks are given in Section \ref{scr}.


\section{Method}\label{sm}
In this section, we discuss the binomial response models with a family of link functions and provide methods of estimating BMD and BMDL using the models.
\subsection{Binomial Response Models}\label{glm}
The Binomial Response Models are members of the Generalized Linear Models (GLMs) described by three components given below.
\begin{enumerate} \item Distributional components: let $y_1,y_2,\ldots,y_n$ be $n$ random samples of adverse events at dose levels $d_1,d_2,\ldots,d_n$, where for each $i\in\{1,2,\ldots,n\}$, $y_i$ has binomial distribution with parameter $(n_i,r_i),\ r_i\in[0,1]$, and $\yb_i=\frac{y_i}{n_i}$ has scaled binomial distribution belongs to the exponential family having the form of probability mass function (pmf) given by \citep{fahrmeirtutz_2001}
\begin{equation*}
s(\yb_i|\theta_i,w_i,\phi)=\exp\left[\frac{\yb_i\theta_i-b(\theta_i)}{\phi}w_i+c(y_i,w_i,\phi)\right],
\end{equation*}
where $r_i=R(d_i)=E(\yb_i|d_i)$, $\theta_i=\log(\frac{r_i}{1-r_i})$ is the so called natural parameters, $b(\theta_i)=\log[1+\exp(\theta_i)]$, $w_i=n_i$, $\phi=1$, and $c(y_i,w_i,\phi)=log\left[\frac{n_i!}{y_i!(n_i-y_i)!}\right]$.

\item Linear predictor: $\eta(d_i)=f(d_i)\betav$, where $f(d_i)$ is a vector function of $d_i$, and $\betav$ is called regression parameter vector.
\item Parametric link function:  $g[\alphav,R(d_i)]=\eta(d_i)$ or $R(d_i)=h\left[\alphav,\eta(d_i)\right]$, where $g$ is called parametric link function and $h$ is the inverse of $g$. We usually assume that the inverse of $g$ exists.

\end{enumerate}
For dose-response studies, the linear predictor is usually assumed as $\eta(d)=\beta_0+\beta_1d$, or $\eta(d)=\beta_0+\beta_1d+\beta_2d^2$, and a single link function such as logistic, probit, log-log, complementary log-log or some other link functions are assumed to fit the models. Here, instead of a single link function, we are using a family of link functions (parametric link function) parameterized by a link parameter vector $\alphav$ to fit the models. So, we are denoting the link function as $g(\alphav,\cdot)$ in place of $g(\cdot)$. \par

Several researchers \citep{1988_stukel,1997_czado} proposed family of link functions (parametric link function) to fit the binomial response models. One such family of link functions for binomial response models is given by
\begin{equation}R(d)=E(\yb|d)=h\left[\alphav,\eta(d)\right]=\frac{\exp\left[G(\alphav,\eta)\right]}{1+\exp\left[G(\alphav,\eta)\right]},\label{fl}
\end{equation}
where $\eta\equiv\eta(d)$, and $G(\alphav,\cdot)$ is called a generating family. There are several forms for Generating family proposed in literature \citep{1988_stukel,1989_czado}. \cite{1988_stukel} provides the following generating family: \\ if $\eta\geq 0$ (i.e., $r\geq \frac12$),
\begin{eqnarray*} G(\alphav,\eta)=\left\{\begin{matrix} \frac{\exp(\alpha_1\eta)-1}{\alpha_1}, & \alpha_1>0\\
                                       \eta, & \alpha_1=0\\
                                       -\frac{\log(1-\alpha_1\eta)}{\alpha_1}, & \alpha_1<0,
                                       \end{matrix}\right.
\end{eqnarray*}
and for $\eta< 0$ (i.e., $r<\frac12$),
\begin{eqnarray*}  G(\alphav,\eta)=\left\{\begin{matrix} \frac{1-\exp(-\alpha_2\eta)}{\alpha_2}, & \alpha_2>0\\
                                       \eta, & \alpha_2=0\\
                                       \frac{\log(1+\alpha_2\eta)}{\alpha_2}, & \alpha_2<0,
                                       \end{matrix}\right.
\end{eqnarray*}
where, $\eta\equiv \eta(d)$, and $r\equiv R(d)$. Note that, for $\alphav=[0,0]'$, we get the logistic link function. So, the logistic link function is a member of this family. Also, several important link functions can be approximated by the members of this family such as Probit link ($\alphav\approx[0.165,0.165]'$), log-log link ($\alphav\approx[-0.037,0.62]'$), and complementary log-log link ($\alphav\approx[0.62,-0.037]'$) \citep{1988_stukel}. \par
For estimating the risk function using the above models, we need to estimate the unknown parameters using a available data sets. Let us denote $\deltav=[\betav',\alphav']'$ for the combined parameter vectors including the unknown regression parameter vector $\betav$ and the link parameter vector $\alphav$. The unknown parameter vector $\deltav$ can be estimated using the Maximum Likelihood Estimation (MLE) methods given in  \cite{1988_stukel}. Due to estimation of the link parameters along with regression parameters, the variances of the estimated regression parameters are increased \citep{1988_taylor}. The variance inflations of the regression parameters are asymptotically zero if the link parameters are orthogonal to the regression parameters \citep{1987_cox}.
\cite{1997_czado} proposed some conditions on the family of link functions providing local orthogonality between link and regression parameter vectors. A family of link functions $\Lambdav=\{h(\alphav,\cdot):\alphav\in\Omegav\}$ provides local orthogonality between link and regression parameter vectors around a point $\eta_0$ asymptotically, if the following conditions are satisfied.
\begin{enumerate} \item There exists $\eta_0$ and $r_0$ such that
\begin{equation}h(\alphav,\eta_0)=r_0,\ \forall\ \alphav\in\Omega, \label{cond1u}\end{equation}
and
\item There exists $s_0$ such that
\begin{equation}\frac{\partial h(\alphav,\eta)}{\partial\eta}\left|_{(\eta=\eta_0)}\right.=s_0,\ \forall\ \alphav\in\Omega,\label{cond2u}\end{equation}
\end{enumerate}
where $\Omegav$ is denoted for the parameter space of $\alphav$.
Such a family $\Lambdav=\{h(\alphav,\cdot):\alphav\in\Omegav\}$ satisfying conditions (\ref{cond1u}) and (\ref{cond2u}) is called $(r_0,s_0)-standardized$ at $\eta_0$ \citep{1997_czado}. \par

Now, for estimating risk function $R(d)$ using $(r_0,s_0)-standardized$ family at $\eta_0$, we need to estimate extra three parameters $r_0,s_0,$ and $\eta_0$.
 For avoiding estimating extra three parameters, \cite{1997_czado} proposed  to choose  $r_0=\beta_0,\ s_0=1$, and $\eta_0=\beta_0$. By choosing the values such a way, the variance inflations of $\betav$ are reduced as the values of $\eta$ vary around the point $\eta_0=\beta_0$, when centered covariates (i.e, $\bar{d}=\frac1n \sum_{i=1}^{n}d_i=0$) are used  \citep{1997_czado}. For this, if dose levels are not centered, we need to transfer the available dose levels as $x_i=d_i-\bar{d}$, and after estimating BMD/BMDL from the model, we make the inverse transformation to get the estimates of BMD/BMDL in the true range of dose levels. For constructing $(r_0=\beta_0,s_0=1)-standardized$ family at $\eta_0=\beta_0$, we adopt the methodologies given by \cite{1997_czado}. Here, we use \cite{1988_stukel}'s generating family to construct the family of link functions, and the $(r_0=\beta_0,s_0=1)-standardized$ at $\eta_0=\beta_0$ generating family is given by:\\
if $\eta_c\geq 0$ [$logit(r)\geq \beta_0$],
\begin{eqnarray} G_c(\alphav,\eta)=\beta_0+\left\{\begin{matrix} \frac{\exp(\alpha_1\eta_c)-1}{\alpha_1}, & \alpha_1>0\\
                                       \eta_c, & \alpha_1=0\\
                                       -\frac{\log(1-\alpha_1\eta_c)}{\alpha_1}, & \alpha_1<0,
                                       \end{matrix}\right.\label{gc1}
\end{eqnarray}
and for $\eta_c< 0$ [$logit(r)<\beta_0$],
\begin{eqnarray}  G_c(\alphav,\eta)=\beta_0+\left\{\begin{matrix} \frac{1-\exp(-\alpha_2\eta_c)}{\alpha_2}, & \alpha_2>0\\
                                       \eta_c, & \alpha_2=0\\
                                       \frac{\log(1+\alpha_2\eta_c)}{\alpha_2}, & \alpha_2<0,
                                       \end{matrix}\right.\label{gc2}
\end{eqnarray}
where $\eta\equiv\eta(d)$, $r\equiv R(d)$, $\eta_{c}=\eta-\beta_0$, and $logit(r)=log[r/(1-r)]$. Hence, the risk function $R(d)$ using the binomial response model with $(r_0=\beta_0,s_0=1)-standardized$ at $\eta_0=\beta_0$ generating family is given by
\begin{equation}R(d)=E(\yb|d)=h\left[\alphav,\eta(d)\right]=\frac{\exp\left[G_c(\alphav,\eta)\right]}{1+\exp\left[G_c(\alphav,\eta)\right]},\label{rd}
\end{equation}
where $\eta\equiv\eta(d)$, and $G_c(\alphav,\cdot)$ is given by equations (\ref{gc1}) \& (\ref{gc2}). In the next section, we provide a expression for the Benchmark dose (BMD) using the above model.

\subsection{Benchmark Dose Estimation}\label{sbmd}
The Benchmark dose (BMD) is defined by the dose level having extra risk $R_E(BMD)=BMR$, where $BMR$ is the Benchmark risk usually pre-specified as $0.01, 0.05$, and $0.1$. So, $BMD$ is the solution of the equation
\begin{eqnarray} R_E(BMD) &=&\frac{R(BMD)-p_0}{1-p_0}= BMR \nonumber\\
\Rightarrow R(BMD)&=&p_0+(1-p_0)BMR
=BMRE, say\nonumber\\
\Rightarrow BMD &=& R^{-1}(BMRE),\label{bmd}
\end{eqnarray}
 where $p_0$ is the Background risk, i.e, the risk at the minimum dose level $d_1$. We denote $\deltav=[\betav',\alphav']'$ for the joint  parameter vector including the regression parameter vector $\betav$, and the link parameter vector $\alphav$. For a fixed value of $BMR\in[0,1]$, the BMD can be expressed as a function of $\deltav$, $S(\deltav)$, say. From equations (\ref{rd}) and (\ref{bmd}), we get a expression for $BMD$ as
\begin{equation}BMD=S(\deltav)=S_1(\deltav)I_{\{LBMR\geq\beta_0\}}+S_2(\deltav)I_{\{LBMR<\beta_0\}},\label{sd}
\end{equation}
where $LBMR=\log(\frac{BMRE}{1-BMRE})$, and $I_{\{LBMR\geq\beta_0\}}$ is the indicator function taking value 1 if $LBMR\geq\beta_0$, and 0 otherwise. The functions $S_1(\deltav)$ and $S_2(\deltav)$  are given by

\begin{eqnarray*}S_1(\deltav)=\left\{\begin{matrix} \frac{\log[\alpha_1(LBMR-\beta_0)+1]}{\alpha_1\beta_1}, & \alpha_1>0\\
                                       \frac{LBMR-\beta_0}{\beta_1}, & \alpha_1=0\\
                                       \frac{1-\exp[-\alpha_1(LBMR-\beta_0)]}{\alpha_1\beta_1}, & \alpha_1<0,
                                       \end{matrix}\right.
\end{eqnarray*}
and,

\begin{eqnarray*} S_2(\deltav)=\left\{\begin{matrix} -\frac{\log[1-\alpha_2(LBMR-\beta_0)]}{\alpha_2\beta_1}, & \alpha_2>0\\
                                       \frac{LBMR-\beta_0}{\beta_1}, & \alpha_2=0\\
                                       \frac{\exp[\alpha_2(LBMR-\beta_0)]-1}{\alpha_2\beta_1}, & \alpha_2<0.
                                       \end{matrix}\right.
\end{eqnarray*}

Note that we require centered dose levels (i.e, $\bar{d}=\frac1n\sum_{i=1}^{n}d_i=0$) for using the model (\ref{rd}). If the dose levels are not centered, we make the transformation $x_i=d_i-\bar{d}$ to have the centered dose levels. After estimating $BMD$ from the model we make the inverse transformation to get the estimated value of BMD within the true range of dose levels.

\subsection{Asymptotic Results}\label{ar}

The asymptotic distributions of unknown parameters for $q$ dimensional multinomial response models with a family of link functions are discussed in \cite{das2014}. For $q=1$, we get the binomial response models using a family of link functions. So, the similar results can be applicable for binomial response models using a family of link functions. However, for making this article self contained, we provide the required asymptotic results here. We denote $\deltavh'=[\betavh',\alphavh']'$ for the MLE of $\deltav=[\betav',\alphav']'$, $l(\deltav)$ for the log-likelihood function, and $\frac{\partial l}{\partial\deltav}$ for the score function for the observed responses. Also, $\Jv_n$ is denoted for the Fisher's information matrix. The asymptotic results are given by the following Lemmas. \par

{\bf Lemma 1: } The score function $\frac{\partial l}{\partial\deltav}$ has an asymptotic multivariate normal distribution with mean $\mathbf{0}$ and variance $\Jv_n$.

{\bf Proof: } From Section \ref{sm}, the risk function is given by,
\begin{equation}R(d)=h[\alphav,\eta(d)],\end{equation}
where $\eta(d)=f(d)\betav$, $\betav$ is an unknown regression parameter vector
  and $\alphav$ is a vector of unknown link parameters.
Also,
\begin{equation}\eta(d)=f(d)\betav=g[\alphav,R(d)],\end{equation}
  where $g$ is the inverse of $h$.

Now, from Section \ref{sm}, the log-likelihood function for the sample
$y_1,\ldots,y_n$ is given by
  \begin{eqnarray}l(\deltav) &=& \sum_{i=1}^n l_i(\deltav)\nonumber\\
  &=& \sum_{i=1}^n[\yb_i\theta_i-b(\theta_i)]n_i+constant.\label{ll}
  \end{eqnarray}
Thus, the score function is \citep[p~436]{fahrmeirtutz_2001},
  \begin{eqnarray}\frac{\partial l(\deltav)}{\partial\deltav} &=& \frac{\partial}{\partial\deltav}\sum_{i=1}^n[\yb_i\theta_i-b(\theta_i)]n_i\nonumber\\
  &=& \sum_{i=1}^n\frac{\partial r_i}{\partial\deltav}[Var(\yb_i)]^{-1}(\yb_i-r_i),
  \label{dlddeltav}
  \end{eqnarray}
  and \citep[p~436]{fahrmeirtutz_2001}
  \begin{eqnarray}-\frac{\partial^2 l(\deltav)}{\partial\deltav\partial\deltav'} &=& \sum_{i=1}^n\frac{\partial r_i}{\partial\deltav}[Var(\yb_i)]^{-1}\frac{\partial r_i}{\partial\deltav'}-
  \sum_{i=1}^n\frac{\partial^2\theta_{i}}{\partial\deltav\partial\deltav'}(\bar{y}_{i}-r_{i})n_i\nonumber\\
  &=& \Hv_n,\ (say).\label{dl2ddeltav}
  \end{eqnarray}
  From equation (\ref{dl2ddeltav}), we get the Fisher information matrix is
  \begin{eqnarray}\Jv_n &=& -E\left[\frac{\partial^2 l(\deltav)}{\partial\deltav\partial\deltav'}\right] 
  = \sum_{i=1}^n\frac{\partial r_i}{\partial\deltav}[Var(\yb_i)]^{-1}\frac{\partial r_i}{\partial\deltav'}.\label{jv}
  \end{eqnarray}

From equation (\ref{dlddeltav}),  using the central limit theorem we have
$\frac{\partial l(\deltav)}{\partial\deltav}$ has asymptotic normal distribution with mean $\mathbf{0}$ and
  variance $\Jv_n$.

  {\bf Lemma 2: } The MLE of $\deltav$, $\deltavh$ has an asymptotic multivariate normal distribution with mean $\deltav$ and variance $\Jv_n^{-1}$.

  {\bf Proof: }   By Taylor series expansion and approximating up to first order term, we have
\begin{eqnarray*}\mathbf{0}=\frac{\partial l(\deltavh)}{\partial\deltav}&=&\frac{\partial l(\deltav)}{\partial\deltav}+
  \left[\frac{\partial^2 l(\deltav)}{\partial\deltav\partial\deltav'}\right](\deltavh-\deltav),
  \end{eqnarray*}
  which gives \citep[p~439]{fahrmeirtutz_2001},
  \begin{eqnarray*}\sqrt{N}(\deltavh-\deltav)&=&
  \sqrt{N}\Hv^{-1}_n\frac{\partial l(\deltav)}{\partial\deltav}\nonumber=\sqrt{N}\Jv_n^{-1}\frac{\partial l(\deltav)}{\partial\deltav}+O_p(N^{-1/2}).\label{tsd}
  \end{eqnarray*}
  Thus, the MLE of $\deltav$, $\deltavh$
  has an asymptotic normal distribution with mean
  $\deltav$ and variance $\Jv_n^{-1}$.

  {\bf Lemma 3: }The estimate $\widehat{BMD}=S(\deltavh)$ is a consistent estimator for $BMD$.

  {\bf Proof: } The proof is trivial from the result that the MLE of $\deltav$, $\deltavh$ is a consistent estimator of $\deltav$, and $S(\deltav)$ is a continuous function of $\deltav$. Hence, $S(\deltavh)$ is a consistent estimator for $S(\deltav)$, i.e., $\widehat{BMD}$ is a consistent estimator for $BMD$.\par

  In the next section, we provide confidence intervals for $BMD$ to find BMDL from the proposed model.

\subsection{Confidence Intervals}\label{sbmdl}

Here, we provide four methods of constructing confidence intervals for $BMD$ for a particular value of $BMR=BMR_0$. The methods are discussed as follows:

\subsubsection{Confidence interval using ML estimates } Here, we use the asymptotic result of the distribution of $\deltavh$ for constructing the confidence interval for BMD. From Lemma 2, we have $\deltavh$ has an asymptotic multivariate normal distribution with mean  $\deltav$ and variance $\Sigma=\Jv_n^{-1}$. Hence, $(\deltavh-\deltav)'\Sigma^{-1}(\deltavh-\deltav)$ has an asymptotic $\chi^2$-distribution with $p$ degrees of freedom, where $p$ is the order of the vector $\deltav$. Hence, the $100(1-\tau)\%$ confidence region for $\deltav$ is given by
\begin{equation}\mathbf{C}=\{\deltav\in\mathbf{R}^p:(\deltavh-\deltav)'\Sigma^{-1}(\deltavh-\deltav)\leq\chi_{p,(1-\tau)}^2\},
\end{equation}
 where $\chi_{p,(1-\tau)}^2$, is the $(1-\tau)$th quantile of the $\chi^2$ distribution with $p$ degrees of freedom. For $BMR=BMR_0$, we compute $BMD=S(\deltav)$, for  $\deltav\in\mathbf{C}$ using equation (\ref{sd}).
 Let us denote
 \begin{eqnarray} S_L&=&Min\{S(\deltav): \deltav\in\mathbf{C}\}, \text{ and  }\nonumber\\
  S_U &=& Max\{S(\deltav): \deltav\in\mathbf{C}\}\label{slu}
 \end{eqnarray}
Now, from (\ref{slu}), we have $\deltav\in\mathbf{C}\Rightarrow S(\deltav)\in[S_L,S_U]$, which implies
$P(S(\deltav)\in[S_L,S_U])\geq P(\deltav\in\mathbf{C})=1-\tau$. Hence, the $100(1-\tau)\%$ conservative confidence interval for $BMD$ is given by $[S_L,S_U]$.

\subsubsection{Confidence interval using LR test } Here, we test the null hypothesis
\begin{eqnarray} H_0: R_E(d)=BMR_0 \text{ vs }     H_1: R_E(d) \neq BMR_0,\label{lrt}
\end{eqnarray}
where $R_E(d)=\frac{R(d)-p_0}{1-p_0}$, with $p_0$ is the background risk.  Let $D(d)$ be the deviance \citep[p~108]{fahrmeirtutz_2001} under null hypothesis and $D(\hat{d})$ be the deviance of the fitted model. Then, $L(d)=D(d)-D(\hat{d})$ has an asymptotic $\chi^2$-distribution with 1 degree of freedom. Let us denote
 \begin{eqnarray} L_{min}&=&Min\{d\in\mathbb{R}: L(d)\leq\chi_{p,(1-\tau)}^2 \}, \text{ and  }\nonumber\\
  L_{max} &=& Max\{d\in\mathbb{R}: L(d)\leq\chi_{p,(1-\tau)}^2 \}\label{lmm}
 \end{eqnarray}
Then, the $100(1-\tau)\%$ confidence interval for $BMD$ is given by $[L_{min},L_{max}]$.

\subsubsection{Confidence interval using score test } Let us denote $u_0=\left[\frac{\partial l}{\partial\beta_0}\right]_{\deltavh_0}$, where $\deltavh_0$ is the MLE of
$\deltav$ under $H_0$ given in equation (\ref{lrt}). Let $\sigmah_0^2$ be the estimated variance
of $u_0$ at $\deltav=\deltavh_0$. Then, $T(d)=u_0^2/\sigmah_0^2$ has an asymptotic $\chi^2$
distribution with $1$ degree of freedom \citep[p~48]{fahrmeirtutz_2001}. Let us denote $T_{min}=min\{d\in\mathbb{R}:T(d)\leq\chi_{1,(1-\tau)}^2\}$, and $T_{max}=max\{d\in\mathbb{R}:T(d)\leq\chi_{1,(1-\tau)}^2\}$. Then, using score test, a $100(1-\tau)\%$ confidence interval for $BMD$ is $[T_{min},T_{max}]$.

Note that the above confidence intervals are by nature two sided. To get one sided confidence interval (BMDL), some researchers \citep{buckley2009confidence,nitcheva2005multiplicity} proposed an adjustment by doubling the significance level of the test and then ignoring the upper limit. So, we construct $100(1-2\tau)\%$ two sided confidence intervals for BMD using the above methods and then the lower limits of that intervals are taken as the one sided $100(1-\tau)\%$ lower confidence bound for BMD.\par

Since, all of the above confidence intervals are constructed using asymptotic results, here we provide a confidence interval using bootstrap technique.

\subsubsection{A bootstrap lower confidence bound } For constructing bootstrap lower confidence bound, we generate $l$ responses $\yv_k=[y_{1k},y_{2k},y_{3k},y_{4k}]'$  using the fitted model $\hat{r}_i=h[\alphavh,\etah(d_i)]$ with $\etah(d_i)=f(d_i)\betavh$, where $y_{ik}$ has binomial distribution with parameter $(n,\hat{r}_i)$ for $i=1,\ldots,4$ and $k=1,\ldots,l$. From the generated $l$ data sets, we estimate BMDs using the proposed method to have samples  $\{BMD_1,BMD_2,\ldots,BMD_{l}\}$ for $BMD$. The $100(1-\tau)\%$ bootstrap lower confidence limit for BMD is given by the $\tau$th quantile of the sample $\{BMD_1,BMD_2,\ldots,BMD_{l}\}$.\par

Let us denote ML, LR, ST, and BT for the methods of estimating BMDL using ML estimates, LR test, score test, and bootstrap technique respectively. Example and simulation studies are provided to illustrate and test the performance of the proposed methods in next sections.

\section{Example: Experiment on Rats Exposed to 1-Bromopropane}\label{ser}
For illustrating the proposed method, we present an example of estimating BMD using a real data set on lung cancer incidence of rats exposed to 1-Bromopropane given in the NTP Technical Report TR-569 \citep{national2011toxicology}. In this study, four groups of rats with each group contains 50 rats are exposed to four dose levels of 1-Bromopropane. After two years of studies, the observed lung cancer incidence of rats at four dose levels 0 ppm , 62.5 ppm, 125 ppm, and 250 ppm  are recorded as $1/50,\ 9/50,\ 8/50,$ and $14/50$ respectively as given in Table \ref{tabrd}.

\begin{table}
  \centering
  \caption{Observed lung cancer incidence of rats exposed to 1-Bromopropane.  }\vspace{0.2 cm}
  \begin{tabular}{cc}
     \hline
     Dose levels ($d_i$) & Responses ($\bar{y}_i$)\\\hline
     0 ppm& 1/50 \\
     62.5 ppm & 9/50 \\
     125 ppm & 8/50\\
     250 ppm & 14/50 \\
     \hline
   \end{tabular}
  \label{tabrd}
\end{table}

\cite{wheeler2012monotonic} also analyzed the same data set noting that ``the data, given in Table \ref{tabrd}, exhibit a linear or supra-linear response indicating that MA may not be able to capture the true D-R relationship". Here, we use the proposed method of using family of link functions (FL) to estimate BMD  and corresponding BMDLs using ML estimates (ML), likelihood ratio test (LR), score test (ST), and bootstrap technique (BT) as described in Sections \ref{sm}. \cite{wheeler2012monotonic} provide the estimates of BMD and corresponding BMDLs using Semi-parametric (Diffuse), Semi-parametric (Historical Controls), Model-averaging, and Quantal-Linear.  In Table \ref{tabrr}, we report the estimated values of BMD using FL and estimated values of BMDLs using four methods ML, LR, ST, and BT within bracket for BMR=0.01, and 0.1. Also, we report the estimated values of BMD and corresponding BMDLs in bracket using Semi-parametric (Diffuse), Semi-parametric (Historical Controls), Model-averaging, and Quantal-Linear for BMR=0.01, and 0.1 from \cite{wheeler2012monotonic} in Table \ref{tabrr}.

\begin{table}
  \small
  \doublespacing
  \centering
  \caption{Estimated values of BMDs and the corresponding BMDLs using Family of link functions, Semi-parametric (Diffuse), Semi-parametric (Historical Controls), Model-averaging, and Quantal-Linear. }\vspace{0.2 cm}
  \begin{tabular}{ccc}
    \hline
  {\multirow{2}{*}{Method}} & \multicolumn{2}{c}{BMR} \\
  \cline{2-3}
    {}   & 0.01 & 0.1\\\hline
    Family of link functions & 8.6 (6.7, 6.2, 6.2, 7.6) & 68.9 (63.5, 50.0, 49.8, 57.6) \\
    Semi-parametric  & \multirow{2}{*}{6.1 (2.1)} & \multirow{2}{*}{56.6 (17.5)}\\
    (Diffuse) & {} & {} \\
    Semi-parametric  & \multirow{2}{*}{6.6 (1.6)} & \multirow{2}{*}{97.1 (23.1)}\\
    (Historical Controls) & {} & {}\\
    Model-averaging & 1.1 (0.14) & 51.1 (17.2)\\
    Quantal-Linear & 7.8 (5.2) & 81.5 (55.0)\\
    \hline
  \end{tabular}
  \label{tabrr}
\end{table}

From Table \ref{tabrr}, we observe that the estimated values of BMD using FL are consistent with the estimated values of BMD using Semi-parametric (Diffuse), Semi-parametric (Historical Controls), and Quantal-Linear for all values of BMR. As mentioned in \cite{wheeler2012monotonic}, the estimated values of BMD using MA diverge and smaller than those using other methods. We observe that the estimated values of BMDLs using FL are higher than those by the methods given in \cite{wheeler2012monotonic}. So, the proposed methods may provide a better estimates of BMDL if the estimated confidence intervals have the expected coverage probabilities for small samples. So, we need to do simulation studies for verifying the coverage probabilities of the proposed confidence intervals for small samples.

\section{Simulation Studies}\label{scma}

In this section, we conduct simulation studies for testing the performance of the proposed methods of estimating BMD and BMDL considering all types of possible cases of generating data sets. Let us denote the proposed method of estimating BMD using the family of link functions as FL. The model averaging method is usually denoted as MA. For testing the performance of FL compare to MA, we provide a simulation study by estimating BMD using FL and MA considering different simulation scenarios with varying sample sizes. Simulation studies are also conducted for testing the performance of the proposed methods of estimating BMDL with respect to their coverage probabilities for small samples.
\subsection{Comparison between FL and MA}\label{scompflma}
We compare the proposed method FL with MA considering the following simulation set up with experimental design consists of four dose levels as $d_1=0,\ d_2=0.25,\ d_3=0.5,$ and $d_4=1.00$ mimicking the design considered by \cite{west2012impact}. Six scenarios for dose response relationships have been considered to  represent all types of possibilities of having probability of adverse events at dose levels varying from shallow to steep curves. The scenarios with true parameter values and the probabilities of adverse events at dose levels are given in Table \ref{tab6s}.

\begin{table}
  \small
  \doublespacing
  \centering
  \caption{Six scenarios for the dose response curves. }\vspace{0.2 cm}
  \begin{tabular}{cccccc}
    \hline
    Scenario & $\deltav=[\beta_0,\beta_1,\alpha_1,\alpha_2]'$ & $R(d_1)$ & $R(d_2)$ & $R(d_3)$ & $R(d_4)$   \\\hline
    1 & $[ -4.5031,4.9075,0.1170,1.5162]'$ & 0.0000 & 0.0015 & 0.0149 & 0.0224 \\
    2 & $[-2.9252,4.9961,1.9078,-1.1403]'$ & 0.0176 & 0.0276 & 0.0760 & 1.0000 \\
    3 & $[-1.3677,2.4678,1.6912,-0.8872]'$ & 0.1067 & 0.1474 & 0.2330 & 0.9856 \\
    4 & $[-0.7784,3.9106,1.6554,-0.8438]'$ & 0.1374 & 0.2060 & 0.3829 & 1.0000 \\
    5 & $[-0.3852,4.7828,1.9908,-0.0870]'$ & 0.0905 & 0.2229 & 0.5058 & 1.0000 \\
    6 & $[1.9190,3.9682,0.9064,0.6930]'$ & 0.1909 & 0.7202 & 0.9000 & 0.9999 \\
    \hline
  \end{tabular}
  \label{tab6s}
\end{table}

 For each scenario, responses ($y_i$) are generated from binomial distribution with parameter $[n,R(d_i)]$, where $n$ is the number of patients administered the dose level $d_i,\ i\in\{1,\ldots,4\}$. For testing the performance of the methods with varying sample sizes and BMR, we consider two different values of $BMR=0.01,\ \&\ 0.1$ and three different sample sizes $n=25,50,100$ for each dose response curve. This provides in total 6 curves $\times$ 2 values of BMR $\times$ 3 sample sizes = 36 different cases. For getting model averaging estimates of BMD, eight standard models \citep{west2012impact} given in Table \ref{tab8m} are considered. The expression for model averaging estimates of BMD, denoted as $\BMDh_{MA}$ is given by
\begin{equation} \BMDh_{MA}=\sum_{k=1}^8 w_k \Bh_k,\end{equation}
where $\Bh_k$ is the estimate of BMD using model $k$, and $w_k=\frac{\exp(-0.5A_k)}{\sum_{k=1}^8\exp(-0.5A_k)}$  with $A_k$ is the Akaike Information Criteria \citep{1973_akaika} given by  $A_k=-2\hat{L}_k+2p_k$, where $\hat{L}_k$ is the maximized log-likelihood value and $p_k$ is the number of parameters in model $k$.

\begin{table}
  \small
  \centering
  \caption{Eight standard models used in MA for computing $\BMDh_{MA}$.}\vspace{0.2 cm}
  \begin{tabular}{ccccc}
    \hline
    Model & Name & R(d) & BMD & Notes \\\hline
    1 & Logistic & $\frac{1}{1+\exp(-\beta_0-\beta_1d)}$ & $\frac{1}{\beta_1}\log\left(\frac{1+e^{-\beta_0}BMR}{1-BMR}\right)$ & None \\\\
    2 & Probit & $\Phi(\beta_0+\beta_1) d$ & $\frac{\Phi^{-1}[BMR(1-\phi_0)+\phi_0]-\beta_0}{\beta_1}$ & {\small $\phi_0=\Phi(\beta_0)$} \\\\
    3 & Quantal-linear & $1-\exp(-\beta_0-\beta_1 d)$ & $\frac{-\log(1-BMR)}{\beta_1}$ & $\beta_0\geq 0,\beta_1\geq 0$\\\\
    4 & Quantal-quadratic & $\gamma_0+(1-\gamma_0)(1-\exp[\beta_1 d^2])$ & $\sqrt{\frac{-\log(1-BMR)}{\beta_1}}$ & $0\leq \gamma_0\leq 1,\beta_1\geq 0$ \\\\
   \multirow{2}{*}{5} & \multirow{2}{*}{Two-stage} & \multirow{2}{*}{$1-\exp(-\beta_0-\beta_1 d-\beta_2 d^2)$} & \multirow{2}{*}{$\frac{-\beta_1+\sqrt{\beta_1^2+4\beta_2T}}{2\beta_2}$} & \multirow{1}{*}{$\beta_j\geq 0,j=0,1,2$} \\
    {} & {} & {} & {} &  {\small $T=-\log(1-BMR)$}\\\\
    \multirow{2}{*}{6} & \multirow{2}{*}{Log-logistic} & \multirow{2}{*}{$\gamma_0+\frac{1-\gamma_0}{1+\exp(-\beta_0-\beta_1\log[d])}$} & \multirow{2}{*}{$\exp\left(\frac{L-\beta_0}{\beta_1}\right)$} & $0\leq\gamma_0\leq 1, \beta_1\geq 0$ \\
    {} & {} & {} & {} & $L=\log(\frac{BMR}{1-BMR})$\\\\
    7 & Log-probit & $\gamma_0+(1-\gamma_0)\Phi[\beta_0+\beta_1\log(x)]$ & $\exp\left[\frac{\Phi^{-1}(BMR)-\beta_0}{\beta_1}\right]$ & $0\leq\gamma_0\leq 1,\beta_1\geq 0$ \\\\
    \multirow{2}{*}{8} & \multirow{2}{*}{Weibull} & \multirow{2}{*}{$\gamma_0+(1-\gamma_0)[1-\exp(-e^{\beta_0}d^{\beta_1})]$} & \multirow{2}{*}{$\exp\left[\frac{\log(T)-\beta_0}{\beta_1}\right]$} & $0\leq\gamma_0\leq 1,\beta_1\geq 0$ \\
    {} & {} & {} & {} & {\small $T=-\log(1-BMR)$}\\
    \hline
  \end{tabular}
  \label{tab8m}
\end{table}
We generate $l=2000$ data sets for each simulation set up. For some cases the simulated responses produce virtually flat dose repones curve \citep{wheeler2009comparing} which does not give any finite estimate of $BMD$. So, we mimic the methodology given in  \cite{wheeler2009comparing} of screening the data sets using Kendall correlation test \citep{kendall1955rank}. We regenerate responses until the responses exhibit the Kendall $p$-value less than or equal to 0.15. \par

 For each cases described above, we estimate BMDs by FL and MA using 2000 simulated data sets. The proposed method FL is compared with MA on the basis of observed absolute relative median bias defined by the absolute value of median$\left[\frac{\hat{BMD}-BMD}{BMD}\right]$ \citep{wheeler2007properties}. The estimated values of $BMD$ are used as a sample of size 2000 for computing absolute relative median bias by FL and MA. Smaller values of absolute relative median bias' are desirable for having better performance by a BMD estimation method. The absolute relative median bias (ARMB) values by FL and MA for each simulation set-up are reported in Table \ref{tabcomp}.
\begin{table}
  \small
  \centering
  \caption{Comparison between FL and MA for accounting model uncertainty in BMD estimation. The observed values of absolute relative median bias' for FL and MA for different cases are reported against the column FL and MA respectively.  }\vspace{0.2 cm}
  \begin{tabular}{ccccccccccc}
    \hline
    Scenario & n & BMR & FL & MA & {} & Scenario & n & BMR & FL & MA \\\hline\\
    \multirow{8}{*}{1} & \multirow{2}{*}{25} & 0.01 & 0.2043 & 0.3149 & {} & \multirow{8}{*}{4} & \multirow{2}{*}{25} & 0.01 & 0.2480 & 7.9541 \\
    {} & {} & 0.1 & 0.0691 & 0.1814 & {} & {} & {} & 0.1 & 0.1420 & 0.8534\\\\
    {} & \multirow{2}{*}{50} & 0.01 & 0.1083 & 0.1384 & {} & {} & \multirow{2}{*}{50} & 0.01 & 0.2925 & 10.1225\\
    {} & {} & 0.1 &  0.0298 & 0.1321 & {} & {} & {} & 0.1 & 0.1563 & 1.2207\\\\
    {} & \multirow{2}{*}{100} & 0.01 & 0.0433 & 0.1392 & {} & {} & \multirow{2}{*}{100} & 0.01 & 0.3092 & 10.4497\\
    {} & {} & 0.1 & 0.0460 & 0.1228 & {} & {} & {} & 0.1 & 0.1834 & 1.2917\\\\

    \multirow{8}{*}{2} & \multirow{2}{*}{25} & 0.01 & 0.0600 & 0.6192 & {} & \multirow{8}{*}{5} & \multirow{2}{*}{25} & 0.01 & 0.5322 & 5.2402 \\
    {} & {} & 0.1 & 0.0520 & 0.0247 & {} & {} & {} & 0.1 & 0.3872 & 0.7904\\\\
    {} & \multirow{2}{*}{50} & 0.01 & 0.0997 & 1.6802 & {} & {} & \multirow{2}{*}{50} & 0.01 & 0.4079 & 7.9412\\
    {} & {} & 0.1 & 0.0212 & 0.4306 & {} & {} & {} & 0.1 & 0.2450 & 1.3338\\\\
    {} & \multirow{2}{*}{100} & 0.01 & 0.0336 & 1.7606 & {} & {} & \multirow{2}{*}{100} & 0.01 & 0.2973 & 11.8273\\
    {} & {} & 0.1 & 0.0193 & 0.4790 & {} & {} & {} & 0.1 & 0.2243 & 2.4060\\\\

    \multirow{8}{*}{3} & \multirow{2}{*}{25} & 0.01 & 0.1059 & 5.7751 & {} & \multirow{8}{*}{6} & \multirow{2}{*}{25} & 0.01 & 0.4644 & 4.3147 \\
    {} & {} & 0.1 & 0.0729 & 0.4777 & {} & {} & {} & 0.1 & 0.3822 & 2.1070\\\\
    {} & \multirow{2}{*}{50} & 0.01 & 0.0773 & 5.8792 & {} & {} & \multirow{2}{*}{50} & 0.01 & 0.2919 & 4.3990\\
    {} & {} & 0.1 & 0.0428 & 0.5366 & {} & {} & {} & 0.1 & 0.2101 & 2.1520\\\\
    {} & \multirow{2}{*}{100} & 0.01 & 0.1487 & 5.4006 & {} & {} & \multirow{2}{*}{100} & 0.01 & 0.1474 & 4.3084\\
    {} & {} & 0.1 & 0.0248 & 0.4850 & {} & {} & {} & 0.1 & 0.1618 & 2.5693\\\\

    \hline
  \end{tabular}
  \label{tabcomp}
\end{table}

From Table \ref{tabcomp}, we see that the observed values of ARMB by FL are very close to those by MA for all values of $n$ and BMR in Scenario 1. So, FL and MA have comparable performance with respect to their observed ARMB values for Scenario 1. Note that the chosen curve for generating data sets in Scenario 1 has very slowly increasing probability of adverse events at dose levels with $R(d_1)=0$, and $R(d_4)=0.0224$. So, it can be concluded that FL and MA provides comparable performance for extremely shallow dose response curves.  If we move towards less shallow dose response curves (Scenarios 2-6), we see that the observed values of ARMB by FL are smaller than those by MA.  For example, in Scenario 4 with $BMR=0.01$, the values of ARMB are 0.5322, 0.4079, \& 0.2973 by method FL, and 5.2402, 7.9412, \& 11.8273 by method MA for sample sizes $n=25,\ 50,$ \& $100$ respectively. Also, for the same Scenario with BMR=0.1, the values of ARMB are 0.3872, 0.2450, \& 0.2243 by method FL, and 0.7904, 1.3338, \& 2.4060 by method MA for sample sizes $n=25,\ 50,$ \& $100$ respectively. This shows that the values of ARMB by FL are smaller than those of ARMB by MA for these cases. Hence, FL performs better than MA with respect to their observed ARMB values for Scenarios 2-6. Also, it is noted that the values of ARMB by MA increase with sample sizes for some scenarios. This shows that the estimates by MA are asymptotically biased when the true models are not included in the model space of MA to estimate BMD.

\subsection{Comparison among Four BMDL Estimation Methods}\label{scs}
Here, we conduct simulation studies to compare four methods of estimating BMDL using ML estimates (ML), likelihood ratio test (LR), score test (ST), and bootstrap technique (BT) with respect to their observed coverage probabilities for small samples. We choose similar simulation set-up considered in Section \ref{scompflma} with the experimental design $\textbf{d}=[0.0,0.25,0.5,1.0]'$ and scenarios given in Table \ref{tab6s} to generate data sets. We also consider two values of $BMR=0.01,$ \& 0.1 and three sample sizes $n=25, 50,$ \& 100 for each scenario. For each simulation set up, we generate $l=1000$ data sets which are also screened  by Kendall correlation test \citep{kendall1955rank} as discussed in Section \ref{scompflma}.\par

 The simulated data sets are used to estimate 95\% BMDL using the four methods ML, LR, ST, and BT. After estimating BMDL using a method, we find an approximate value of coverage probability given by $\frac{N_l}{l}$, where $N_{l}$ is the number of times the estimated values of BMDL are less than or equal to BMD out of $l$ data sets generated. The coverage probabilities by four methods ML, LR, ST, and BT for each simulation set-up are reported in Table \ref{tabcovp}.\par

 \begin{table}
  \small
  \centering
  \caption{Comparison among four methods of estimating BMDL with respect to their coverage probabilities for different simulation set-up. The observed coverage probabilities of ML, LR, ST, and BT are given against the column ML, LR, ST, and BT respectively.}\vspace{0.2 cm}
  \begin{tabular}{ccccccccccccccc}
    \hline
    \multirow{2}{*}{Scenario} & \multirow{2}{*}{n} & \multirow{2}{*}{BMR} & \multicolumn{4}{c}{Methods} & {} & \multirow{2}{*}{Scenario} & \multirow{2}{*}{n} & \multirow{2}{*}{BMR} & \multicolumn{4}{c}{Methods} \\\cline{4-7}\cline{12-15}
    {} & {} & {} & ML & LR & ST & BT & {} & {} & {} & {} & ML & LR & ST & BT \\\hline\\
    \multirow{8}{*}{1} & \multirow{2}{*}{25} & 0.01 & 1.00 & 1.00 & 1.00 & 1.00 & {} & \multirow{8}{*}{4} & \multirow{2}{*}{25} & 0.01 &  1.00 & 1.00 & 1.00 & 1.00 \\
    {} & {} & 0.1 &  1.00 & 1.00 & 1.00 & 1.00 & {} & {} & {} & 0.1 &  1.00 & 1.00 & 1.00 & 1.00\\\\
    {} & \multirow{2}{*}{50} & 0.01 &  1.00 & 1.00 & 1.00 & 1.00 & {} & {} & \multirow{2}{*}{50} & 0.01 &  1.00 & 1.00 & 1.00 & 1.00\\
    {} & {} & 0.1 &   1.00 & 1.00 & 1.00 & 1.00 & {} & {} & {} & 0.1 &  1.00 & 1.00 & 1.00 & 1.00\\\\
    {} & \multirow{2}{*}{100} & 0.01 &  1.00 & 1.00 & 1.00 & 1.00 & {} & {} & \multirow{2}{*}{100} & 0.01 &  1.00 & 1.00 & 1.00 & 1.00\\
    {} & {} & 0.1 &  1.00 & 1.00 & 1.00 & 0.95 & {} & {} & {} & 0.1 &  0.99 & 1.00 & 1.00 & 1.00\\\\

     \multirow{8}{*}{2} & \multirow{2}{*}{25} & 0.01 & 1.00 & 1.00 & 1.00 & 0.68 & {} & \multirow{8}{*}{5} & \multirow{2}{*}{25} & 0.01 &  1.00 & 0.99 & 1.00 & 0.88 \\
    {} & {} & 0.1 &  1.00 & 1.00 & 1.00 & 1.00 & {} & {} & {} & 0.1 &  0.97 & 0.99 & 1.00 & 1.00\\\\
    {} & \multirow{2}{*}{50} & 0.01 &  1.00 & 1.00 & 1.00 & 0.91 & {} & {} & \multirow{2}{*}{50} & 0.01 &  1.00 & 0.99 & 1.00 & 0.93\\
    {} & {} & 0.1 &   1.00 & 1.00 & 1.00 & 1.00 & {} & {} & {} & 0.1 &  0.94 & 0.99 & 1.00 & 1.00\\\\
    {} & \multirow{2}{*}{100} & 0.01 &  0.99 & 1.00 & 1.00 & 0.98 & {} & {} & \multirow{2}{*}{100} & 0.01 &  1.00 & 0.98 & 1.00 & 0.95\\
    {} & {} & 0.1 &  1.00 & 0.99 & 1.00 & 1.00 & {} & {} & {} & 0.1 &  0.91 & 0.97 & 1.00 & 1.00\\\\

     \multirow{8}{*}{3} & \multirow{2}{*}{25} & 0.01 & 1.00 & 1.00 & 1.00 & 1.00 & {} & \multirow{8}{*}{6} & \multirow{2}{*}{25} & 0.01 &  1.00 & 0.99 & 0.98 & 0.76 \\
    {} & {} & 0.1 &  1.00 & 1.00 & 1.00 & 1.00 & {} & {} & {} & 0.1 &  1.00 & 1.00 & 1.00 & 1.00\\\\
    {} & \multirow{2}{*}{50} & 0.01 &  1.00 & 1.00 & 1.00 & 1.00 & {} & {} & \multirow{2}{*}{50} & 0.01 &  1.00 & 1.00 & 1.00 & 0.77\\
    {} & {} & 0.1 &   1.00 & 1.00 & 1.00 & 1.00 & {} & {} & {} & 0.1 &  1.00 & 1.00 & 1.00 & 1.00\\\\
    {} & \multirow{2}{*}{100} & 0.01 &  1.00 & 1.00 & 1.00 & 1.00 & {} & {} & \multirow{2}{*}{100} & 0.01 &  1.00 & 1.00 & 1.00 & 0.83\\
    {} & {} & 0.1 &  1.00 & 1.00 & 1.00 & 1.00 & {} & {} & {} & 0.1 &  1.00 & 1.00 & 1.00 & 1.00\\\\

    \hline
  \end{tabular}
  \label{tabcovp}
\end{table}

From Table \ref{tabcovp}, we see that the observed coverage probabilities by LR and ST are greater than 0.95 for all scenarios and BMR values with all sample sizes. The method BT fails to provide the expected coverage probabilities for Scenario 2 with $n=25,50$ and Scenarios 5 \& 6 for all sample sizes when BMR=0.01. The observed coverage probabilities by ML also exceed the expected probability 0.95 for all the cases except for Scenario 5 with $n=50,100$, when BMR=0.1. We also studied the observed average length of one sided confidence interval (average of [BMD-$\hat{BMDL}$]) by four methods. We observe that BT provides smallest values and ML \& ST provide largest values for the average of (BMD-$\hat{BMDL}$) for all the cases considered. Hence, we conclude that LR is best among all the methods of estimating BMDL with respect to coverage probabilities and lengths of the confidence intervals.

\section{Conclusions}\label{scr}

For accounting model uncertainty in BMD estimation, a family of link functions for binary response models are used to develop a method for estimating BMD. The family of link functions provides local orthogonality between link and regression parameters to reduce the variance inflations of the estimated regression parameters. Infinite number of link functions including some standard link functions are the members of this family. For accounting model uncertainty in BMD estimation, the family of link functions provides a better approach than model averaging method as the model space considered in MA to get model averaged estimate usually contains only a finite number of models. Methods of estimating BMDL are also provided using the family of link functions. \par

The proposed method is illustrated by an example with a real data set observing that FL is consistent with the existing results in literature. By comparing FL with MA using simulation studies considering different simulation scenarios, we see that FL outperforms MA for most of the scenarios. Simulation studies are also conducted to compare the four methods of estimating BMDL and we see that LR is best among the four methods of estimating BMDL considering both the coverage probability as well as the length of the confidence intervals.  \par

There are other methods exist in literature using Bayesian and non parametric approach for accounting model uncertainty in BMD estimation. The frequentist methods are usually easy to implement and require less time for computations than other non frequentist approach. We compared FL with MA as both the methods are based on frequentist approach to deal with the model uncertainty problems. In future, the proposed method may be compared with other non frequentist approach to estimate BMD to test the performance of FL.

\newpage
\bibliographystyle{elsarticle-harv}
\bibliography{reidview}

\begin{thebibliography}{24}
\expandafter\ifx\csname natexlab\endcsname\relax\def\natexlab#1{#1}\fi
\expandafter\ifx\csname url\endcsname\relax
  \def\url#1{\texttt{#1}}\fi
\expandafter\ifx\csname urlprefix\endcsname\relax\def\urlprefix{URL }\fi

\bibitem[{Akaike(1973)}]{1973_akaika}
Akaike, H., 1973. Information theory and an extension of the maximum likelihood
  principle. In: Petrov, B.~N., Csaki, B. (Eds.), Proceedings of the Second
  International Symposium on Information Theory. Akademiai Kiado, Budapest, pp.
  267--281.

\bibitem[{Bailer et~al.(2005)Bailer, Noble, and Wheeler}]{2005_bailer}
Bailer, A.~J., Noble, R.~B., Wheeler, M.~W., 2005. Model uncertainty and risk
  estimation for experimental studies of quantal responses. Risk Analysis
  25~(2), 291--299.
\newline\urlprefix\url{http://dx.doi.org/10.1111/j.1539-6924.2005.00590.x}

\bibitem[{Buckley et~al.(2009)Buckley, Piegorsch, and
  West}]{buckley2009confidence}
Buckley, B.~E., Piegorsch, W.~W., West, R.~W., 2009. Confidence limits on
  one-stage model parameters in benchmark risk assessment. Environmental and
  ecological statistics 16~(1), 53--62.

\bibitem[{Cox and Reid(1987)}]{1987_cox}
Cox, D.~R., Reid, N., 1987. Parameter orthogonality and approximate conditional
  inference. Journal of the Royal Statistical Society 49, 1--39.

\bibitem[{Crump(1984)}]{1984_crump}
Crump, K.~S., 1984. A new method for determining allowable daily intakes.
  Fundamental and applied toxicology 4~(5), 854--871.

\bibitem[{Czado(1989)}]{1989_czado}
Czado, C., 1989. Link misspecification and data selected transformations in
  binary regression models. Tech. rep., Ph.D. Thesis. School of Operations
  Research and Industrial Engineering, Cornell University, Ithaca, NY.

\bibitem[{Czado(1997)}]{1997_czado}
Czado, C., 1997. On selecting parametric link transformation families in
  generalized linear models. Journal of Statistical Planning and inference 61,
  125--139.

\bibitem[{Das and Mukhopadhyay(2014)}]{das2014}
Das, I., Mukhopadhyay, S., 2014. On generalized multinomial models and joint
  percentile estimation. Journal of Statistical Planning and Inference 145,
  190--203.

\bibitem[{Fahrmeir and Tutz(2001)}]{fahrmeirtutz_2001}
Fahrmeir, L., Tutz, G., 2001. Multivariate Statistical Modelling Based on
  Generalized Linear Models, 2nd Edition. Springer, New York.

\bibitem[{Kang et~al.(2000)Kang, Kodell, and Chen}]{kang2000incorporating}
Kang, S.-H., Kodell, R.~L., Chen, J.~J., 2000. Incorporating model
  uncertainties along with data uncertainties in microbial risk assessment.
  Regulatory Toxicology and Pharmacology 32~(1), 68--72.

\bibitem[{Kendall(1955)}]{kendall1955rank}
Kendall, M.~G., 1955. Rank correlation methods. Hafner Publishing Co, New York.

\bibitem[{Morales et~al.(2006)Morales, Ibrahim, Chen, and
  Ryan}]{morales2006bayesian}
Morales, K.~H., Ibrahim, J.~G., Chen, C.-J., Ryan, L.~M., 2006. Bayesian model
  averaging with applications to benchmark dose estimation for arsenic in
  drinking water. Journal of the American Statistical Association 101~(473),
  9--17.

\bibitem[{Nitcheva et~al.(2005)Nitcheva, Piegorsch, Webster~West, and
  Kodell}]{nitcheva2005multiplicity}
Nitcheva, D.~K., Piegorsch, W.~W., Webster~West, R., Kodell, R.~L., 2005.
  Multiplicity-adjusted inferences in risk assessment: Benchmark analysis with
  quantal response data. Biometrics 61~(1), 277--286.

\bibitem[{Piegorsch et~al.(2013)Piegorsch, An, Wickens, Webster~West, Pe{\~n}a,
  and Wu}]{piegorsch2013information}
Piegorsch, W.~W., An, L., Wickens, A.~A., Webster~West, R., Pe{\~n}a, E.~A.,
  Wu, W., 2013. Information-theoretic model-averaged benchmark dose analysis in
  environmental risk assessment. Environmetrics 24~(3), 143--157.

\bibitem[{Program et~al.(2011)}]{national2011toxicology}
Program, N.~T., et~al., 2011. Toxicology and carcinogenesis studies of
  1-bromopropane (cas no. 106-94-5) in f344/n rats and b6c3f1 mice (inhalation
  studies). National Toxicology Program technical report series~(564), 1.

\bibitem[{Shao and Small(2011)}]{shao2011potential}
Shao, K., Small, M.~J., 2011. Potential uncertainty reduction in model-averaged
  benchmark dose estimates informed by an additional dose study. Risk Analysis
  31~(10), 1561--1575.

\bibitem[{Shao and Small(2012)}]{shao2012statistical}
Shao, K., Small, M.~J., 2012. Statistical evaluation of toxicological
  experimental design for bayesian model averaged benchmark dose estimation
  with dichotomous data. Human and Ecological Risk Assessment: An International
  Journal 18~(5), 1096--1119.

\bibitem[{Simmons et~al.(2015)Simmons, Chen, Li, Wang, Piegorsch, Fang, Hu, and
  Dunn}]{2015_simmons}
Simmons, S.~J., Chen, C., Li, X., Wang, Y., Piegorsch, W.~W., Fang, Q., Hu, B.,
  Dunn, G.~E., 2015. Bayesian model averaging for benchmark dose estimation.
  Environmental and Ecological Statistics 22~(1), 5--16.
\newline\urlprefix\url{http://dx.doi.org/10.1007/s10651-014-0285-4}

\bibitem[{Stukel(1988)}]{1988_stukel}
Stukel, T.~A., 1988. Generalized logistic models. Journal of the American
  Statistical Association 83, 426--431.

\bibitem[{Taylor(1988)}]{1988_taylor}
Taylor, J. M.~G., 1988. The cost of generalized logistic regression. Journal of
  the American Statistical Association 83, 1078--1083.

\bibitem[{West et~al.(2012)West, Piegorsch, Pe{\~n}a, An, Wu, Wickens, Xiong,
  and Chen}]{west2012impact}
West, R.~W., Piegorsch, W.~W., Pe{\~n}a, E.~A., An, L., Wu, W., Wickens, A.~A.,
  Xiong, H., Chen, W., 2012. The impact of model uncertainty on benchmark dose
  estimation. Environmetrics 23~(8), 706--716.

\bibitem[{Wheeler and Bailer(2012)}]{wheeler2012monotonic}
Wheeler, M., Bailer, A.~J., 2012. Monotonic bayesian semiparametric benchmark
  dose analysis. Risk Analysis 32~(7), 1207--1218.

\bibitem[{Wheeler and Bailer(2007)}]{wheeler2007properties}
Wheeler, M.~W., Bailer, A.~J., 2007. Properties of model-averaged bmdls: a
  study of model averaging in dichotomous response risk estimation. Risk
  Analysis 27~(3), 659--670.

\bibitem[{Wheeler and Bailer(2009)}]{wheeler2009comparing}
Wheeler, M.~W., Bailer, A.~J., 2009. Comparing model averaging with other model
  selection strategies for benchmark dose estimation. Environmental and
  Ecological Statistics 16~(1), 37--51.

\end{thebibliography}

\end{document}